\documentclass[11pt, a4paper]{article}

\usepackage[a4paper, top=2.5cm, bottom=2.5cm, left=2cm, right=2cm]{geometry}

\usepackage[english]{babel}
\usepackage[T1]{fontenc}
\usepackage{lmodern}

\usepackage{amsmath}
\usepackage{authblk}
\usepackage{hyperref}
\usepackage{setspace}
\usepackage{booktabs}
\usepackage{enumitem}
\usepackage{tabularx}
\usepackage{array}

\usepackage{amsmath}
\usepackage{authblk}
\usepackage{hyperref}
\usepackage{setspace}
\usepackage{booktabs} 
\usepackage{enumitem} 

\hypersetup{
    colorlinks=true,
    linkcolor=blue,
    filecolor=magenta,      
    urlcolor=cyan,
    citecolor=blue,
}

\title{\textbf{How Indian Dermatologists are Utilizing Artificial Intelligence for Clinical Practice and Workflow Management: A Nationwide Survey with a Special Focus on atopic dermatitis}}
\author[1]{Dipayan Sengupta}
\author[2]{Saumya Panda}
\author[3]{Sandipan Dhar}
\author[4]{Dipankar De}
\author[5]{Deepika Pandhi}
\author[6]{Narayanan B}

\affil[1]{Charnock Hospital, Kolkata, India}
\affil[2]{Department of Dermatology, Jagannath Gupta Institute of Medical Sciences and Hospital, Kolkata, India}
\affil[3]{Department of Pediatric Dermatology, Institute of Child Health, Kolkata-700017, India}
\affil[4]{Department of Dermatology, Venereology, and Leprology, Postgraduate Institute of Medical Education and Research, Chandigarh-160012, India}
\affil[5]{Department of Dermatology and STD, University College of Medical Sciences \& GTBH, University of Delhi, Delhi-110095, India}
\affil[6]{Department of DVL, Sree Balaji Medical College and Hospital, Chennai, India}

\date{\today}

\begin{document}

\maketitle

\begin{abstract}
\noindent \textbf{Background:} The development of artificial intelligence (AI) in dermatology has historically prioritized visual diagnostic algorithms, while other clinical and administrative components of chronic disease management have received comparatively less attention. This study utilized a problem-first methodology to map the clinical bottlenecks reported by Indian dermatologists---with a specific focus on atopic dermatitis (AD)---against their current adoption of AI tools.

\vspace{0.2cm}
\noindent \textbf{Methods:} In a study commissioned by the Society for Eczema Studies (SES), a nationwide, cross-sectional survey was administered to practicing Indian dermatologists ($N = 377$). The instrument evaluated technology-agnostic clinical frustrations and AD-specific management hurdles prior to assessing AI usage patterns, technological barriers, and ethical apprehensions. Data were analyzed using descriptive statistics, Pearson’s Chi-square ($\chi^2$) tests, Benjamini-Hochberg False Discovery Rate (FDR) corrections, and multivariate logistic regression.

\vspace{0.2cm}
\noindent \textbf{Results:} Chronic disease management challenges, including patient adherence (61.3\%) and treatment planning in difficult or refractory cases (57.0\%), were reported more frequently than diagnostic uncertainty (48.0\%). In AD management, objective severity scoring (e.g., EASI/SCORAD) was commonly reported as a challenge (47.7\%) and had the lowest satisfaction rate among the measured workflow areas. Current AI adoption was reported by 49.9\% of respondents and was most commonly represented by general Large Language Models (LLMs), used for tasks such as summarizing research and drafting clinical or academic content rather than specialized image analysis.

Barriers to adoption varied by experience level: veteran dermatologists ($>20$ years practice) more frequently cited a lack of training (64.3\%, $p = 0.019$), whereas junior dermatologists ($\le5$ years) more frequently cited a lack of clinical utility after trying AI tools (22.8\%, $p = 0.038$). AI users were also more likely than non-users to report concern regarding patient self-misdiagnosis and anxiety. After adjustment for clinical experience and academic affiliation, active AI usage remained independently associated with this concern (adjusted odds ratio [aOR]: 2.25, 95\% CI: 1.45 -- 3.50, $p = 0.0003$). AI users also expressed higher concern regarding use by non-dermatologists after FDR correction ($p_{adj} = 0.0251$).

\vspace{0.2cm}
\noindent \textbf{Conclusion:} Indian dermatologists reported frequent use of general-purpose AI tools, particularly for literature synthesis, documentation, and academic tasks, while their highest reported clinical needs related to chronic disease management and AD workflow support. AI users were more likely than non-users to report concerns regarding patient self-diagnosis and non-specialist use. Future dermatology AI tools may be more clinically useful if developed around clinician-supervised workflow support rather than standalone diagnostic applications.
\end{abstract}

\newpage
\doublespacing

\section{Introduction}

\subsection{The ``Technology-Push'' Dilemma in Dermatological AI}
Dermatology, an inherently visual and pattern-driven specialty, has been at the forefront of the artificial intelligence (AI) revolution in medicine. Over the past decade, the translation of clinical acumen into computational algorithms---primarily via Convolutional Neural Networks (CNNs)---has demonstrated notable mathematical proficiency. Seminal studies have established that AI systems can achieve diagnostic accuracies comparable to human experts in controlled, \textit{in-silico} environments, particularly in dermato-oncology for the binary classification of melanoma versus benign pigmented lesions [1]. 

However, despite these algorithmic advances, the successful translation and integration of AI into routine clinical workflows remain limited. A significant barrier to clinical adoption is the prevailing ``technology-push'' model of development. AI solutions in dermatology have largely been engineered based on what the technology \textit{can} do (e.g., image classification of discrete, high-acuity lesions) rather than what practicing dermatologists actually \textit{need} them to do. This has created a literature landscape that views AI primarily as a ``diagnostic oracle,'' often giving less emphasis to the broader spectrum of daily dermatological practice, including administrative burdens, patient counselling, and longitudinal disease management.

\subsection{The Need for a ``Problem-First'' Framework}
To bridge the gap between technological capability and clinical utility, there is a need to apply a ``problem-first'' or customer development framework---borrowed from lean startup methodologies---to medical informatics. Before developing or evaluating a Clinical Decision Support System (CDSS), it is important to understand the ``job-to-be-done'' [2]. This requires mapping the foundational, technology-agnostic clinical frictions that dermatologists face daily, as well as analyzing the current habits and workarounds they employ to solve these problems in the absence of AI.

While several international surveys have recently explored dermatologists' attitudes toward AI---including studies across Australia [3], Saudi Arabia [4, 5], Germany [6], and China [7]---the majority of these assessments follow a traditional format. They typically measure general enthusiasm, perceived threats (such as deskilling or the ``black box'' phenomenon), and baseline familiarity [8]. However, they rarely utilize a dynamic, problem-centric approach that isolates a clinician's specific workflow frustrations \textit{prior} to introducing the concept of AI. Consequently, the existing literature has had limited ability to identify the intersection between a felt clinical need and a viable technological solution. 

While recent international surveys of dermatologists [3, 8] have effectively measured general attitudes toward AI, they largely assess these perceptions in isolation. However, broader clinical informatics frameworks, such as the NASSS framework [9], indicate that technological integration may fail when developers do not adequately understand foundational, technology-agnostic clinical friction. Therefore, similar to landmark workflow analyses in primary care [10], there is value in mapping the actual `jobs-to-be-done' and current problem-solving habits of dermatologists before proposing algorithmic solutions [11].

\subsection{The Chronic Care Complexity: Atopic Dermatitis as a Paradigm}
The limitations of the current ``diagnostic oracle'' paradigm are particularly relevant to the management of chronic inflammatory skin diseases, such as atopic dermatitis (AD). Unlike skin cancer, where the primary clinical endpoint may be an excision or biopsy, AD management is longitudinal and complex [12]. 

The daily reality of AD care involves a high cognitive and communicative load: assessing objective severity through time-consuming scoring systems (e.g., EASI, SCORAD), navigating nuanced therapeutic step-ups (e.g., the transition to biologics or JAK inhibitors), managing multi-system comorbidities, and ensuring long-term patient adherence amidst caregiver anxiety and topical steroid phobia [13]. Furthermore, the recent proliferation of Generative AI, Large Language Models (LLMs), and emerging Large Vision-Language Models (LVLMs, such as MedGemma) presents new multimodal capabilities that may assist with some of these communicative and administrative burdens [14]. Yet, chronic inflammatory conditions remain underrepresented in AI dermatology literature, leaving an evidence gap in our understanding of how modern technology might alleviate bottlenecks in chronic disease management.

\subsection{Identification of the Knowledge Gap and Study Rationale}
A synthesis of the current literature reveals three interconnected knowledge gaps:
\begin{enumerate}
    \item \textbf{The Absence of Needs-Based Mapping:} There is a lack of structured research that isolates and validates the daily, non-technological frustrations of dermatologists before assessing their AI usage, limiting developers' understanding of the baseline clinical habits AI must improve upon.
    \item \textbf{The Chronic Disease Void:} The literature heavily favors acute image diagnostics, while giving less attention to the potential of AI, particularly Generative AI, in the longitudinal and administrative management of complex chronic conditions like atopic dermatitis.
    \item \textbf{Geographical and Demographic Underrepresentation:} There is limited data regarding AI adoption and clinical friction among dermatologists in the Indian subcontinent---a demographic managing high patient volumes and a diversity of skin types.
\end{enumerate}

To address these gaps, this nationwide, cross-sectional study was designed using a ``customer development'' methodology. By deploying a dynamically branching survey instrument, this study systematically maps the core clinical and administrative challenges of Indian dermatologists---with a specialized focus on AD management---and cross-references these reported problems against their current habits and adoption of AI tools. Ultimately, this study aims to transition the discourse from theoretical AI capabilities to real-world, unmet clinical needs, providing data to guide the future development of targeted, workflow-integrated dermatological technologies.

\section{Materials and Methods}

\subsection{Study Design and Setting}
A nationwide, cross-sectional observational survey, commissioned by the Society for Eczema Studies (SES) was conducted to evaluate the clinical challenges and artificial intelligence (AI) adoption patterns among dermatologists practicing in India. The study was conducted between June 15, 2025, and April 23, 2026. The survey specifically targeted board-certified dermatologists and dermatology residents; participation was strictly restricted to practitioners currently residing and working within India.

\subsection{Sample Size Determination}
The target population was defined using the approximate number of dermatologists in India (using the membership count of the Indian Association of Dermatologists, Venereologists and Leprologists (IADVL), which comprised roughly 17,000 registered dermatologists at the commencement of the study). To achieve a 95\% confidence level with a 5\% margin of error and an assumed response distribution of 50\% (to yield the most conservative sample size), the required sample size was calculated using the finite population correction formula:

\begin{equation}
n = \frac{N \cdot Z^2 \cdot p(1-p)}{(N-1)e^2 + Z^2 \cdot p(1-p)}
\end{equation}

Where $N = 17,000$, $Z = 1.96$ (for 95\% confidence), $p = 0.5$, and $e = 0.05$. This calculation yielded a target minimum sample size of 376 respondents.

\subsection{Survey Instrument Development and Validation}
The survey instrument was developed through a multi-step process. Initially, a core group of dermatologists and health informaticians drafted a comprehensive questionnaire focusing on general dermatology practice bottlenecks, atopic dermatitis (AD) specific challenges, and AI utilization. Prior to widespread distribution, a formal validation study was conducted among a localized cohort. This pilot phase assessed face validity, content validity, and the cognitive load required to complete the questionnaire. The final instrument comprised modules on: (1) baseline demographics, (2) clinical friction points, (3) AI tool usage and utility, and (4) societal and ethical apprehensions. The complete, dynamically branching survey instrument is provided in \textbf{Supplementary Appendix S4}.

\subsection{Recruitment and Data Collection}
Eligible participants were invited through a multimodal digital outreach strategy via the Society for Eczema Studies (SES) and associated professional digital portals. Based on digital analytics across the professional portals and social media networks utilized, the estimated survey reach was approximately 4,000 to 5,000 dermatologists, yielding an estimated response rate of approximately 8\% ($N=377$). To ensure data integrity, respondent IP addresses were temporarily utilized exclusively by a backend protocol for duplicate-prevention deduplication. These identifiers were immediately hashed and discarded prior to data analysis to ensure complete respondent anonymity in compliance with local data protection standards.

\subsection{Ethical Considerations}
The study was conducted in strict accordance with the ethical principles outlined in the Declaration of Helsinki. Explicit digital informed consent was obtained from all participants prior to accessing the survey questions. Even though this was a voluntary, anonymous, non-interventional survey assessing physician workflows and collected no patient-level Protected Health Information (PHI), formal Institutional Ethics Committee (IEC) approval was sought and obtained (Ref No: JIMSH/IEC-2026/05-01).

\subsection{Statistical Analysis}
Data were analyzed utilizing Python (version 3.10) using \texttt{pandas}, \texttt{statsmodels}, and \texttt{scipy.stats}. Bivariate associations were assessed using Pearson’s Chi-square ($\chi^2$) tests with Yates' continuity correction. To control for Type I errors across multiple hypothesis testing regarding ethical concerns, a Benjamini-Hochberg False Discovery Rate (FDR) correction was applied. To evaluate whether the observed association between AI use and heightened concern regarding patient self-misdiagnosis and anxiety was confounded by clinical experience or academic affiliation, a multivariate logistic regression was performed adjusting for these variables. A two-tailed adjusted $p$-value of $\le 0.05$ was considered statistically significant.

\section{Results}

\subsection{Respondent Demographics and Practice Characteristics}
A total of 377 dermatologists completed the survey. The majority of respondents were engaged in clinical practice ($n = 290$, 76.9\%), while 39.5\% ($n = 149$) held academic or research roles. Respondents could select multiple roles, indicating overlap between clinical and academic positions.

Respondents represented a wide range of clinical experience: 40.1\% ($n = 151$) had less than 5 years of post-residency experience, and 29.4\% ($n = 111$) had more than 20 years of experience. The most frequently reported practice settings were private clinics (40.1\%) and government medical colleges (34.5\%). 

To validate the cohort's relevance to the atopic dermatitis (AD) specific modules, respondents were asked to estimate their AD patient volume. Of the 356 respondents who completed this module (21 respondents bypassed this specific section due to incomplete submissions or lack of routine AD exposure), the majority (62.1\%) reported that AD patients constitute greater than 10\% of their total clinical workload (Table 1).

\begin{table}[htbp]
\centering
\caption{Baseline Demographics and Practice Characteristics ($N = 377$)}
\label{tab:demographics}
\begin{tabular}{lc}
\toprule
\textbf{Characteristic} & \textbf{n (\%)} \\
\midrule
\textbf{Current Role}\textsuperscript{a} & \\
\quad Dermatologist (Clinical Practice) & 290 (76.9) \\
\quad Dermatologist (Academic/Research) & 149 (39.5) \\
\addlinespace
\textbf{Years of Practice (Post-Residency)} & \\
\quad Less than 5 years & 151 (40.1) \\
\quad 5 -- 10 years & 59 (15.6) \\
\quad 11 -- 20 years & 56 (14.9) \\
\quad More than 20 years & 111 (29.4) \\
\addlinespace
\textbf{Place of Practice}\textsuperscript{a} & \\
\quad Own Clinic & 151 (40.1) \\
\quad Government Medical College & 130 (34.5) \\
\quad Private Medical College & 96 (25.5) \\
\quad Corporate Hospital & 44 (11.7) \\
\quad Polyclinic & 42 (11.1) \\
\quad Government Hospital & 16 (4.2) \\
\addlinespace
\textbf{Proportion of atopic dermatitis Patients}\textsuperscript{b} & \textbf{($n = 356$)} \\
\quad 0 -- 10\% & 135 (37.9) \\
\quad 11 -- 25\% & 169 (47.5) \\
\quad 26 -- 50\% & 40 (11.2) \\
\quad More than 50\% & 12 (3.4) \\
\bottomrule
\end{tabular}
\vspace{0.1cm} \\
\parbox{\textwidth}{\small \textsuperscript{a} Multiple responses permitted; column percentages sum to $>100\%$. \\ \textsuperscript{b} Denominator is 356 due to missing responses ($n=21$). Percentages calculated based on valid responses.}
\end{table}

\subsection{Clinical Challenges: General vs. Atopic Dermatitis}
Respondents identified managing long-term patient adherence ($n = 231$, 61.3\%) and treatment planning in difficult or refractory cases ($n = 215$, 57.0\%) as the most frequent general clinical challenges. Socioeconomic and psychological factors, including the cost of therapy (47.7\%) and psychological impact on the patient (44.8\%), were also reported. Diagnostic uncertainty in complex cases was reported by 48.0\% ($n = 181$) of the cohort.

Consistent with the broader challenges of chronic care, management of refractory AD was the most frequently selected disease-specific hurdle ($n = 209$, 55.4\%). AD score calculation (e.g., EASI/SCORAD) was also commonly reported as a clinical challenge (47.7\%). Difficulty in diagnosis for unusual presentations of AD was additionally reported by 36.1\% ($n = 136$) of the respondents (Table 2).

\begin{table}[htbp]
\centering
\caption{Frequency of Reported Clinical Challenges ($N = 377$)}
\label{tab:challenges}
\begin{tabular}{lc}
\toprule
\textbf{Challenge Category} & \textbf{n (\%)} \\
\midrule
\textbf{General Dermatological Challenges} & \\
\quad Managing long-term patient adherence & 231 (61.3) \\
\quad Treatment planning in difficult/refractory cases & 215 (57.0) \\
\quad Diagnostic uncertainty in complex cases & 181 (48.0) \\
\quad Cost of therapy & 180 (47.7) \\
\quad Psychological impact on the patient & 169 (44.8) \\
\quad Checking for drug interactions & 98 (26.0) \\
\quad Image analysis (Histopathology/Dermatoscopy) & 94 (24.9) \\
\quad Keeping up with new research/treatments & 94 (24.9) \\
\addlinespace
\textbf{atopic dermatitis Specific Challenges} & \\
\quad Managing refractory / difficult cases & 209 (55.4) \\
\quad Score calculation (e.g., EASI/SCORAD) & 180 (47.7) \\
\quad Patient/caregiver counselling and lifestyle advice & 151 (40.1) \\
\quad Difficulty in diagnosis when presentation is unusual & 136 (36.1) \\
\bottomrule
\end{tabular}
\vspace{0.1cm} \\
\parbox{\textwidth}{\small \textit{Note:} Respondents could select multiple challenges. Percentages reflect the proportion of the total cohort.}
\end{table}

\subsection{Current Workarounds, Satisfaction, and Opportunity Mapping}
For respondents who selected specific workflow challenges, the survey also assessed current methods used to address these problems and satisfaction with those methods. Satisfaction was defined as the proportion of respondents reporting that they were either ``Satisfied'' or ``Very Satisfied'' with their current approach. This analysis was available for seven workflow areas for which matched current-method and satisfaction items were included in the survey instrument.

The lowest overall satisfaction rate was observed for AD score calculation (58.9\%), despite 47.7\% of the total cohort selecting it as a challenge. Within this group, many respondents reported either manual calculation, visual estimation, or not routinely performing formal scoring. Diagnostic uncertainty also showed a comparatively lower satisfaction rate (68.5\%), with respondents commonly relying on senior consultation, online search, reference tools, and professional groups. By contrast, image analysis had the highest satisfaction rate among the measured domains (76.6\%) and was selected by a smaller proportion of respondents (24.9\%) (Table 3).

\begin{table}[htbp]
\centering
\caption{Reported Workflow Challenges and Satisfaction with Current Methods}
\label{tab:opportunity}
\small
\begin{tabularx}{\textwidth}{>{\raggedright\arraybackslash}X
                                >{\centering\arraybackslash}p{3cm}
                                >{\centering\arraybackslash}p{2.5cm}
                                >{\raggedright\arraybackslash}X}
\toprule
\textbf{Problem Area} & \textbf{Respondents Selecting Problem} & \textbf{Satisfaction Rate\textsuperscript{a}} & \textbf{Interpretation} \\
\midrule
AD score calculation & 180 (47.7\%) & 58.9\% & Lowest satisfaction; manual or omitted scoring common \\
Diagnostic uncertainty & 181 (48.0\%) & 68.5\% & Reliance on senior consultation, search, and groups \\
AD refractory management & 209 (55.4\%) & 72.2\% & Common challenge with expert- and literature-dependent workflows \\
AD diagnosis difficulty & 136 (36.1\%) & 75.0\% & Existing methods relatively satisfactory \\
Treatment planning & 215 (57.0\%) & 75.3\% & High-frequency need with moderate-to-high satisfaction \\
AD patient counselling & 151 (40.1\%) & 75.5\% & Verbal counselling predominates; satisfaction relatively high \\
Image analysis & 94 (24.9\%) & 76.6\% & Lower frequency and highest satisfaction among measured domains \\
\bottomrule
\end{tabularx}
\vspace{0.1cm} \\
\parbox{\textwidth}{\small \textsuperscript{a} Satisfaction rate defined as the proportion reporting ``Satisfied'' or ``Very Satisfied'' among respondents who selected that problem.}
\end{table}

Overall, the opportunity mapping suggested that the most relevant areas for future AI or clinical decision support development may not be determined by frequency alone. AD score calculation, diagnostic uncertainty, and refractory AD management combined relatively high reported frequency with lower satisfaction or reliance on time-intensive workarounds. In contrast, image analysis, although a major historical focus of dermatology AI research, was less frequently selected and had relatively high satisfaction with existing approaches in this cohort.

\subsection{AI Adoption, Usage Patterns, and Practice Variables}
Current use of AI-powered tools or software in professional workflows was reported by 49.9\% ($n = 188$) of respondents. Conversely, 47.5\% ($n = 179$) reported non-use, and 2.7\% ($n = 10$) indicated that such tools were not applicable to their practice. Among the 188 users, general Large Language Models (LLMs) (e.g., ChatGPT, Gemini) were the most frequently utilized tool type. Reported primary use cases included summarizing research and medical literature, assisting with differential diagnosis, and academic preparation.

AI adoption rates varied by clinical role. Dermatologists with academic or research involvement reported higher AI usage (58.4\%) compared to those in strictly clinical practice (44.3\%) ($\chi^2 = 6.61, p = 0.010$).

Among the 179 non-users, reported barriers to adoption varied significantly according to years of clinical experience. Non-users with more than 20 years of experience were more likely to cite a ``lack of training or unsure how to use them'' compared to those with 20 years or less of experience (64.3\% vs. 43.9\%, $p = 0.019$). Non-users with 5 years or less of experience were more likely to report having ``tried a few but were not impressed'' compared to non-users with greater than 5 years of experience (22.8\% vs. 9.7\%, $p = 0.038$) (Table 4).

\begin{table}[htbp]
\centering
\caption{Factors Associated with AI Adoption and Barriers}
\label{tab:factors}
\begin{tabular}{lccc}
\toprule
\textbf{Variable} & \textbf{AI Users (\%)} & \textbf{Non-Users (\%)} & \textbf{$p$-value\textsuperscript{*}} \\
\midrule
\textbf{Adoption by Current Role} & & & \textbf{0.010} \\
\quad Academic/Research Involved ($n=149$) & 87 (58.4) & 62 (41.6) & \\
\quad Strictly Clinical Practice ($n=228$) & 101 (44.3) & 127 (55.7) & \\
\midrule
\textbf{Barrier: ``Lack of Training''\textsuperscript{†}} & \textbf{Selected} & \textbf{Not Selected} & \textbf{0.019} \\
\quad $>$ 20 years practice & 36 (64.3) & 20 (35.7) & \\
\quad $\le$ 20 years practice & 54 (43.9) & 69 (56.1) & \\
\midrule
\textbf{Barrier: ``Tried but not impressed''\textsuperscript{†}} & \textbf{Selected} & \textbf{Not Selected} & \textbf{0.038} \\
\quad $\le$ 5 years practice & 13 (22.8) & 44 (77.2) & \\
\quad $>$ 5 years practice & 12 (9.7) & 110 (90.3) & \\
\bottomrule
\end{tabular}
\vspace{0.1cm} \\
\parbox{\textwidth}{\small \textsuperscript{*} $p$-values calculated using Pearson’s Chi-square test with Yates' continuity correction. Significant values ($p \le 0.05$) are bolded. \\ \textsuperscript{†} Analyzed exclusively within the subgroup of Non-AI Users ($n = 179$).}
\end{table}

\subsection{Apprehensions Regarding AI Integration}
Overall, the potential for misdiagnosis or errors ($n = 240$, 65.4\%) and the risk of deskilling or over-reliance on AI ($n = 172$, 46.9\%) were the most frequently reported concerns regarding the integration of AI in dermatology.

We evaluated whether specific ethical apprehensions differed between AI users and non-users after correction for multiple testing. Following Benjamini-Hochberg FDR correction, the elevated concern regarding patient anxiety and self-misdiagnosis ($p_{adj} = 0.0015$) and non-derm usage ($p_{adj} = 0.0251$) remained statistically significant among AI users compared to non-users. Conversely, the concern regarding algorithmic bias was reduced to a non-significant trend ($p_{adj} = 0.0743$). 

Furthermore, a multivariate logistic regression was performed to assess whether the association between AI usage and concern regarding patient self-misdiagnosis/anxiety was explained by clinical experience or academic affiliation. Even after controlling for years of clinical experience and academic affiliation, active AI usage remained a significant independent predictor of concern regarding patient self-medication and anxiety (adjusted odds ratio [aOR]: 2.25, 95\% CI: 1.45 -- 3.50, $p = 0.0003$). This suggests that the observed association was not fully explained by these measured demographic or professional variables.

There were no statistically significant differences between AI users and non-users regarding direct clinical concerns over the potential for misdiagnosis or errors (64.4\% vs. 66.5\%, $p = 0.695$) or deskilling and over-reliance on AI (47.3\% vs. 46.4\%, $p = 0.975$) (Table 5).

\begin{table}[htbp]
\centering
\caption{Frequency of Specific AI Concerns Stratified by Usage}
\label{tab:concerns}
\small
\begin{tabularx}{\textwidth}{>{\raggedright\arraybackslash}X
                                >{\centering\arraybackslash}p{2.8cm}
                                >{\centering\arraybackslash}p{2.8cm}
                                >{\centering\arraybackslash}p{2.2cm}}
\toprule
\textbf{Primary Concern} & \textbf{AI Users ($n=188$)} & \textbf{Non-Users ($n=179$)} & \textbf{Raw $p$-value\textsuperscript{*}} \\
\midrule
\textbf{Societal \& Ethical Concerns} & & & \\
\quad Patient self-misdiagnosis \& anxiety & 137 (72.9\%) & 97 (54.2\%) & \textbf{$<0.001$} \\
\quad Non-Derm Usage / Increased competition & 104 (55.3\%) & 74 (41.3\%) & \textbf{0.010} \\
\quad Algorithmic bias (e.g., in skin tones) & 92 (48.9\%) & 68 (38.0\%) & \textbf{0.044} \\
\addlinespace
\textbf{Direct Clinical Concerns} & & & \\
\quad Potential for misdiagnosis or errors & 121 (64.4\%) & 119 (66.5\%) & 0.751 \\
\quad Deskilling or over-reliance on AI & 89 (47.3\%) & 83 (46.4\%) & 0.934 \\
\bottomrule
\end{tabularx}
\vspace{0.1cm} \\
\parbox{\textwidth}{\small \textsuperscript{*} $p$-values calculated using Pearson’s Chi-square test with Yates' continuity correction prior to FDR adjustment. Significant raw values ($p \le 0.05$) are bolded.}
\end{table}

\section{Discussion}

The clinical integration of artificial intelligence into dermatology has long been shaped by a focus on image-based diagnosis of cutaneous malignancies [1, 15]. However, by employing a problem-first, customer-development methodology, this cross-sectional survey of 377 Indian dermatologists suggests a potential mismatch between a major area of historical technological development and some currently reported clinical needs. Our findings indicate that when dermatologists' foundational workflow frustrations are isolated, frequently reported challenges relate to chronic disease management, treatment planning, adherence, and administrative workload, alongside diagnostic uncertainty.

\subsection{Opportunity Mapping: From Reported Need to Workflow Support}
Prior to assessing technology usage, our study mapped the daily ``jobs-to-be-done'' within the clinic. Respondents most frequently identified patient adherence (61.3\%) and treatment planning for difficult or refractory cases (57.0\%) as clinical hurdles, followed by diagnostic uncertainty (48.0\%), cost of therapy (47.7\%), AD score calculation (47.7\%), and psychological impact on the patient (44.8\%).

However, the opportunity mapping analysis suggests that frequency alone may not identify the highest-yield targets for AI or clinical decision support. AD score calculation emerged as an important workflow gap because it combined a high reported frequency with the lowest satisfaction rate among measured domains. Many respondents relied on manual calculation, visual estimation, or did not routinely perform formal scoring. This pattern suggests that current approaches to objective severity measurement may not be well aligned with the pace of routine clinical practice.

Diagnostic uncertainty and refractory AD management also appear to be relevant areas for clinician-facing support. These domains were commonly reported and were addressed through methods such as senior consultation, online search, reference tools, professional groups, and literature review. These approaches are familiar and often useful, but may be time-intensive or dependent on local access to expertise. In contrast, image analysis was selected by fewer respondents and had the highest satisfaction rate among measured domains. Although image-based diagnosis remains an important area of dermatology AI, our findings suggest that surveyed clinicians perceived substantial needs in broader workflow domains, including severity scoring, refractory disease management, diagnostic uncertainty, and patient counselling.

When cross-referenced with actual tool adoption, a shift in utility is evident. While nearly half of the cohort (49.9\%) reported using AI tools, adoption was largely represented by general Large Language Models (LLMs) rather than specialized dermatological image analysis software. This adoption pattern is consistent with recent literature suggesting a role for generative AI as a cognitive support tool [14]. Rather than primarily seeking an algorithm for visual diagnosis, clinicians in this cohort reported using LLMs for tasks such as summarizing research for complex treatment planning, drafting clinical notes, and structuring academic work. This indicates that future digital health interventions in dermatology may achieve higher clinical uptake by integrating into the cognitive and administrative workflows of chronic disease management, rather than focusing exclusively on primary visual identification [16].

\subsection{Atopic Dermatitis and the Demand for Lower-Friction Severity Quantification}
By focusing specifically on atopic dermatitis, this study highlights the workflow burden associated with chronic inflammatory care. Formal scoring systems (e.g., EASI, SCORAD) are standard prerequisites for initiating advanced systemic therapies such as biologics or JAK inhibitors [12], but their routine use can be time-consuming. In our cohort, AD score calculation was reported as a challenge by 47.7\% of respondents and had the lowest satisfaction rate among the measured workflow areas.

This finding suggests that current ``app-based'' medical tools, which often require manual data entry, may not fully address the practical constraints of busy clinical encounters [17]. Based on these findings, there may be an unmet need for lower-friction severity quantification tools. Such tools could include streamlined calculators, structured electronic templates, or computer-vision-assisted systems that help estimate severity from clinical images while preserving clinician oversight [18].

\subsection{AI Users and Heightened Concern Regarding Downstream Risks}
While previous international surveys [3, 7, 8] have identified generalized fears of AI, our data showed a specific pattern in the perception of risk. General concern regarding algorithmic accuracy, including misdiagnosis or errors, was similar between AI users and non-users. However, active AI users were more likely to report concerns about patient self-misdiagnosis/anxiety and use by non-dermatologists.

We use the term ``Adopter's Paradox'' to describe this pattern: respondents who used AI were also more likely to report concern about its potential misuse by patients or non-specialists. This should be interpreted as an association rather than evidence of causality. Specifically, 72.9\% of AI users expressed concern over patient self-misdiagnosis and anxiety, compared to 54.2\% of non-users, and this association remained significant after adjustment for clinical experience and academic affiliation. In the context of AD---a predominantly pediatric condition where caregiver anxiety and topical corticosteroid phobia are recognized barriers to care [13]---this concern may have practical relevance. Clinicians utilizing LLMs may recognize the authoritative and conversational tone of these models [19]. Consequently, they may be more aware that unfiltered, consumer-facing AI could contribute to misinformation, self-medication, or disruption of clinician-mediated care [20].

AI users also expressed higher concern regarding algorithmic bias, although this was reduced to a non-significant trend after FDR correction ($p_{adj} = 0.0743$). This concern is consistent with recent dermatological dataset audits, which have reported underrepresentation of diverse Fitzpatrick skin tones in public training data [21]. AI users were also significantly more likely to report concern regarding ``non-dermatologist usage'' (55.3\% vs. 41.3\%, $p_{adj} = 0.0251$), highlighting perceived risks related to the use of AI-assisted tools by non-specialists.

\subsection{The Generational Divide: Redefining Technology Acceptance}
By deconstructing the barriers among the non-adopter subgroup, this study refines monolithic models of technology acceptance in healthcare [22]. Resistance to AI integration was stratified by clinical experience, representing distinct archetypes of friction.

Among veteran dermatologists ($>20$ years of practice), the primary barrier was structural: 64.3\% cited a ``lack of training or unsure how to use them'' ($p = 0.019$). Conversely, early-career dermatologists ($\le5$ years of practice) were significantly more likely to cite an experiential utility barrier, reporting that they had ``tried a few but were not impressed'' (22.8\%, $p = 0.038$). This generational divide implies that veteran clinicians may require intuitive, low-friction interfaces and educational support to overcome onboarding hurdles. In contrast, early-career clinicians may require tools with clearer clinical utility and stronger evidence of benefit in real-world workflows.

\subsection{Limitations}
This study contains several methodological limitations. First, the reliance on digital recruitment channels (webinars, professional portals, social media groups) and non-probability convenience sampling introduces inherent selection bias. The sample likely overrepresents tech-literate, digitally engaged dermatologists, potentially inflating baseline AI awareness estimates. Second, based on the estimated digital reach, the response rate was approximately 8\%; thus, national representativeness relative to the entire dermatologist community of India cannot be definitively established. Third, while the survey instrument underwent expert face and content validation during a pilot phase, exhaustive psychometric testing (e.g., test-retest reliability) was not performed. Finally, the cross-sectional nature of the data precludes causal inferences. Regarding the observed association between AI usage and heightened concern regarding patient self-diagnosis, we cannot definitively conclude whether AI usage \textit{causes} heightened ethical concern, or if clinicians who are inherently more cautious or observant are more likely to utilize these tools.

\section{Conclusion}
By applying a problem-first methodology, this study suggests that AI integration in dermatology may be occurring in ways that differ from the traditional emphasis on diagnostic image analysis. Indian dermatologists in this cohort reported using LLMs for cognitive and administrative tasks associated with complex, chronic diseases such as atopic dermatitis. At the same time, AI users were more likely to report concerns regarding patient self-diagnosis and use by non-specialists. Future technological development may be more clinically useful if it emphasizes clinician-supervised, workflow-integrated tools that address reported bottlenecks such as AD severity scoring, refractory disease management, diagnostic uncertainty, and patient education while maintaining patient safety and specialist oversight.

\section*{Acknowledgments and Declarations}
\noindent \textbf{Study Commissioning:} This study was formally commissioned by the Society for Eczema Studies (SES), India, to identify technological and clinical gaps in the management of atopic dermatitis and broader dermatological workflows. \\
\textbf{Conflict of Interest:} The authors declare no conflicts of interest.

\section*{References}
\begin{enumerate}[label={[\arabic*]}, leftmargin=*, itemsep=0.1cm]
\item Esteva A, Kuprel B, Novoa RA, Ko J, Swetter SM, Blau HM, Thrun S. Dermatologist-level classification of skin cancer with deep neural networks. \textit{Nature}. 2017;542(7639):115-118.
\item Christensen CM, Hall T, Dillon K, Duncan DS. Know your customers' ``jobs to be done''. \textit{Harvard Business Review}. 2016;94(9):54-62.
\item Partridge B, Janda M, Gillespie N, Silva CV, Arnold C, Abbott L, et al. Attitudes towards the use of artificial intelligence in dermatology: a survey of Australian dermatologists. \textit{Australas J Dermatol}. 2025;66(5):e279-e286.
\item Al-Ali F, Polesie S, Paoli J, Aljasser M, Salah LA. Attitudes towards artificial intelligence among dermatologists working in Saudi Arabia. \textit{Dermatol Pract Concept}. 2023;13(1):e2023035.
\item Esmaeel SE, Alruwailiy RF, Alanazi ZM, Alanazi AA, Aldhabyan FN, Fahmy EK. Perceptions of healthcare providers on artificial intelligence integration into dermatology clinical practice; a cross-sectional study in the Kingdom of Saudi Arabia. \textit{Int J Med Dev Ctries}. 2024;8(6):1478.
\item Augustin M, Reinders P, Janke TM, Strömer K, von Kiedrowski R, Kirsten N, et al. Attitudes toward and use of eHealth technologies among German dermatologists: repeated cross-sectional survey in 2019 and 2021. \textit{J Med Internet Res}. 2024;26:e45817.
\item Shen C, Li C, Xu F, Wang Z, Shen X, Gao J, et al. Web-based study on Chinese dermatologists’ attitudes towards artificial intelligence. \textit{Ann Transl Med}. 2020;8(11):698.
\item Polesie S, Gillstedt M, Kittler H, Lallas A, Tschandl P, Zalaudek I, Paoli J. Attitudes towards artificial intelligence within dermatology: an international online survey. \textit{Br J Dermatol}. 2020;183(1):159-161.
\item Greenhalgh T, Wherton J, Papoutsi C, Lynch J, Hughes G, A'Court C, et al. Beyond adoption: a new framework for theorizing and evaluating nonadoption, abandonment, and challenges to the scale-up, spread, and sustainability of health and care technologies. \textit{J Med Internet Res}. 2017;19(11):e367.
\item Sinsky C, Colligan L, Li L, Prgomet M, Reynolds S, Goeders L, et al. Allocation of physician time in ambulatory practice: a time and motion study in 4 specialties. \textit{Ann Intern Med}. 2016;165(11):753-760.
\item Yang Q, Steinfeld A, Rosé C, Zimmerman J. Unremarkable AI: fitting intelligent decision support into critical, clinical decision-making processes. \textit{Proc CHI Conf Hum Factors Comput Syst}. 2019:1-11.
\item Wollenberg A, Kinberger M, Arents B, Aszodi N, Avila Valle G, Barbarot S, et al. European guideline (EuroGuiDerm) on atopic eczema: part I–systemic therapy. \textit{J Eur Acad Dermatol Venereol}. 2022;36(9):1409-1431.
\item Li AW, Yin ES, Antaya RJ. Topical corticosteroid phobia in atopic dermatitis: a systematic review. \textit{JAMA Dermatol}. 2017;153(10):1036-1042.
\item Thirunavukarasu AJ, Ting DSJ, Elangovan K, Gutierrez L, Ting TF, Ting DSW. Large language models in medicine. \textit{Nat Med}. 2023;29(8):1930-1940.
\item Gomolin A, Netchiporouk E, Gniadecki R, Litvinov IV. Artificial intelligence applications in dermatology: where do we stand? \textit{Front Med (Lausanne)}. 2020;7:100.
\item Kelly CJ, Karthikesalingam A, Suleyman M, Corrado G, King D. Key challenges for delivering clinical impact with artificial intelligence. \textit{BMC Med}. 2019;17(1):1-9.
\item Chopra R, Vakharia PP, Sacotte R, Silverberg JI. Severity strata for Eczema Area and Severity Index (EASI), modified EASI, Scoring Atopic Dermatitis (SCORAD), objective SCORAD, Atopic Dermatitis Severity Index and body surface area in adolescents and adults with atopic dermatitis. \textit{Br J Dermatol}. 2017;177(5):1316-1321.
\item Coiera E, Koch S. The shifting sands of ambient intelligence. \textit{Yearb Med Inform}. 2022;31(1):12-16.
\item Ayers JW, Poliak A, Dredze M, Leas EC, Zhu Z, Kelley JB, et al. Comparing physician and artificial intelligence chatbot responses to patient questions posted to a public social media forum. \textit{JAMA Intern Med}. 2023;183(6):589-596.
\item Haupt CE, Marks M. AI-generated medical advice—GPT and beyond. \textit{JAMA}. 2023;329(16):1351-1352.
\item Daneshjou R, Smith MP, Sun MD, Rotemberg V, Zou J. Lack of transparency and potential bias in artificial intelligence data sets and algorithms: a scoping review. \textit{JAMA Dermatol}. 2021;157(11):1362-1369.
\item Holden RJ, Karsh BT. The technology acceptance model: its past and its future in health care. \textit{J Biomed Inform}. 2010;43(1):159-172.
\end{enumerate}

\newpage
\section*{Supplementary Appendix}

\subsection*{S1. Analysis of Non-AI Workflows and Satisfaction Metrics}
To understand the baseline clinical friction prior to AI intervention, respondents were asked to identify their methods for solving specific clinical challenges and rate their satisfaction with those methods on a 4-point Likert scale (Dissatisfied, Neutral, Satisfied, Very Satisfied). 

The data suggest a discrepancy in satisfaction between different workflow areas, with the lowest satisfaction observed for atopic dermatitis (AD) severity scoring.

\subsubsection*{S1.1. Comparative Satisfaction Across Workflow Areas}
Satisfaction was defined as the proportion of respondents reporting either ``Satisfied'' or ``Very Satisfied'' among those who selected the corresponding problem.

\begin{table}[htbp]
\centering
\caption{Satisfaction with Current Methods by Reported Problem Area}
\label{tab:supp_satisfaction}
\begin{tabular}{lcc}
\toprule
\textbf{Problem Area} & \textbf{Respondents Selecting Problem} & \textbf{Satisfied or Very Satisfied} \\
\midrule
Image analysis & 94 & 76.6\% \\
AD patient counselling & 151 & 75.5\% \\
Treatment planning & 215 & 75.3\% \\
AD diagnosis difficulty & 136 & 75.0\% \\
AD refractory management & 209 & 72.2\% \\
Diagnostic uncertainty & 181 & 68.5\% \\
AD score calculation & 180 & 58.9\% \\
\bottomrule
\end{tabular}
\end{table}

\subsubsection*{S1.2. Atopic Dermatitis Severity Scoring (e.g., EASI, SCORAD)}
The calculation of objective severity scores demonstrated the lowest satisfaction rate among the workflow areas measured in the survey.

\vspace{0.2cm}
\noindent \textbf{Methods Used for AD Scoring ($n=180$ respondents who selected this problem; multiple selections permitted)}
\begin{itemize}[itemsep=0cm]
    \item Manual Calculation (Pen/Paper): $n=73$ (40.6\%)
    \item Not routinely performed in clinical practice: $n=69$ (38.3\%)
    \item Visual Estimation (No formal scoring): $n=53$ (29.4\%)
    \item Web/App-based Calculator: $n=52$ (28.9\%)
    \item Research setting only: $n=37$ (20.6\%)
\end{itemize}

\vspace{0.2cm}
\noindent \textbf{Satisfaction with Current Scoring Methods ($n=180$)}
\begin{itemize}[itemsep=0cm]
    \item Satisfied or Very Satisfied: 58.9\%
    \item Neutral or Dissatisfied: 41.1\%
\end{itemize}

\noindent \textit{Insight:} The relatively frequent use of manual calculation, visual estimation, and omission of formal scoring suggests that current tools may not fully meet the needs of routine clinical practice.

\subsubsection*{S1.3. Patient Counselling and Lifestyle Modification (AD)}
Despite the time-consuming nature of patient education in chronic diseases, the majority of dermatologists relied on verbal communication.

\vspace{0.2cm}
\noindent \textbf{Methods Used for AD Counselling ($n=151$ respondents who selected this problem; multiple selections permitted)}
\begin{itemize}[itemsep=0cm]
    \item Verbal Discussion: $n=141$ (93.4\%)
    \item Printed Handouts: $n=30$ (19.9\%)
    \item Directing to Web Resources: $n=19$ (12.6\%)
    \item Referral to a Counsellor: $n=11$ (7.3\%)
\end{itemize}

\vspace{0.2cm}
\noindent \textbf{Satisfaction with Current Counselling Methods ($n=151$)}
\begin{itemize}[itemsep=0cm]
    \item Satisfied or Very Satisfied: 75.5\%
    \item Neutral or Dissatisfied: 24.5\%
\end{itemize}

\noindent \textit{Insight:} The reliance on verbal discussion and relatively high satisfaction suggests that AI tools designed for patient education may be most useful if they support clinician-led counselling rather than replacing it with generic chatbot interfaces.

\subsubsection*{S1.4. General Diagnostic Uncertainty}
Traditional methods for resolving diagnostic uncertainty yielded lower satisfaction than several other measured domains, but higher satisfaction than AD scoring.

\vspace{0.2cm}
\noindent \textbf{Methods Used for Diagnostic Uncertainty ($n=181$ respondents who selected this problem; multiple selections permitted)}
\begin{itemize}[itemsep=0cm]
    \item Consult Seniors / Colleagues: $n=130$ (71.8\%)
    \item Online Search: $n=105$ (58.0\%)
    \item Use Standard Reference Tools / Books: $n=94$ (51.9\%)
    \item Post in Professional Groups: $n=85$ (47.0\%)
    \item Refer Patient: $n=20$ (11.0\%)
\end{itemize}

\vspace{0.2cm}
\noindent \textbf{Satisfaction with Diagnostic Resolution Methods ($n=181$)}
\begin{itemize}[itemsep=0cm]
    \item Satisfied or Very Satisfied: 68.5\%
    \item Neutral or Dissatisfied: 31.5\%
\end{itemize}

\subsubsection*{S1.5. Treatment Planning and Refractory AD Management}
Treatment planning and refractory AD management were among the most frequently selected challenge areas. Current approaches were generally associated with moderate-to-high satisfaction, although they often depended on guidelines, literature search, and senior consultation.

\vspace{0.2cm}
\noindent \textbf{Treatment Planning ($n=215$ respondents who selected this problem; multiple selections permitted)}
\begin{itemize}[itemsep=0cm]
    \item Review Guidelines: $n=178$ (82.8\%)
    \item Consult Seniors: $n=125$ (58.1\%)
    \item Use Reference Tools: $n=96$ (44.7\%)
    \item Trial and Error: $n=90$ (41.9\%)
    \item Satisfied or Very Satisfied: 75.3\%
\end{itemize}

\vspace{0.2cm}
\noindent \textbf{AD Refractory Management ($n=209$ respondents who selected this problem; multiple selections permitted)}
\begin{itemize}[itemsep=0cm]
    \item Literature Search: $n=153$ (73.2\%)
    \item Consult Seniors: $n=128$ (61.2\%)
    \item Trial and Error: $n=70$ (33.5\%)
    \item Refer Higher Center: $n=48$ (23.0\%)
    \item Satisfied or Very Satisfied: 72.2\%
\end{itemize}

\subsection*{S2. Additional Cross-Tabulations}

\subsubsection*{S2.1. Practice Setting vs. AI Utility}
While the main text highlights the academic vs. clinical divide, an analysis of practice settings (Strictly Government vs. Strictly Private/Mixed) reveals differing utility profiles among AI users. 

\begin{itemize}[itemsep=0cm]
    \item \textbf{Drafting Clinical Notes:} Private/Mixed practice dermatologists were significantly more likely to use AI for drafting clinical notes compared to Government dermatologists (58.3\% vs. 19.3\%, $p = 0.013$).
    \item \textbf{Patient Education:} Similarly, Private/Mixed practitioners utilized AI for generating patient education materials at nearly triple the rate of Government practitioners (75.0\% vs. 26.3\%, $p = 0.005$).
\end{itemize}

\noindent \textit{Insight:} This may reflect different operational realities; private clinics may have more opportunity to dedicate resources to personalized patient communication and detailed documentation, whereas government settings may be constrained by higher patient volumes.

\subsection*{S3. Full Survey Dataset}
The full result dataset utilized for this analysis will be published in the final published version only.

\subsection*{S4. Full Survey Instrument}
\noindent \textbf{Study Title:} Mapping Challenges and AI Adoption in Dermatology with a Special Focus on atopic dermatitis \\
\textbf{Format:} Digital/Online Questionnaire \\
\textbf{Estimated Completion Time:} 5--8 minutes

\subsubsection*{Section 1: Informed Consent}
\noindent \textbf{Q. Digital Informed Consent:} \textit{By clicking ``Agreed'' below, you acknowledge that you have read the study information, you are a registered dermatologist or dermatology resident practicing in India, and you voluntarily consent to participate in this anonymous survey. No patient-level data or personally identifiable information will be collected.}
\begin{itemize}[label={[ ]}, itemsep=0cm]
    \item Agreed \textit{(Proceeds to survey)}
    \item Declined \textit{(Terminates survey)}
\end{itemize}

\subsubsection*{Section 2: Baseline Demographics}
\noindent \textbf{Q1. What is your current primary role? (Select all that apply)}
\begin{itemize}[label={[ ]}, itemsep=0cm]
    \item Dermatologist (Clinical Practice)
    \item Dermatologist (Academic/Research)
    \item Dermatology Resident / Post-Graduate
    \item Other (Please specify)
\end{itemize}

\noindent \textbf{Q2. What is your primary place of practice? (Select all that apply)}
\begin{itemize}[label={[ ]}, itemsep=0cm]
    \item Own Clinic
    \item Polyclinic
    \item Corporate Hospital
    \item Private Medical College
    \item Govt Medical College
    \item Govt Hospital
\end{itemize}

\noindent \textbf{Q3. How many years of clinical experience do you have?}
\begin{itemize}[label={[ ]}, itemsep=0cm]
    \item Less than 5 years
    \item 5 -- 10 years
    \item 11 -- 20 years
    \item More than 20 years
\end{itemize}

\noindent \textbf{Q4. What is your estimated volume of atopic dermatitis (AD) patients in your routine practice?}
\begin{itemize}[label={[ ]}, itemsep=0cm]
    \item 0 -- 10\%
    \item 11 -- 25\%
    \item 26 -- 50\%
    \item More than 50\%
\end{itemize}

\subsubsection*{Section 3: Clinical Frictions \& Workflow}
\noindent \textbf{Q5. What are the primary day-to-day clinical challenges you face in your practice? (Select all that apply)}
\begin{itemize}[label={[ ]}, itemsep=0cm]
    \item Diagnostic Uncertainty
    \item Treatment Planning
    \item Image Analysis / Dermoscopy Interpretation
    \item Patient Adherence
    \item Keeping Up-to-Date with Literature
    \item Cost of Therapy
    \item Psychological Impact of Disease
    \item Drug Interactions
    \item AD Score Calculation (e.g., EASI, SCORAD)
    \item AD Patient Counselling
    \item AD Refractory Management
    \item AD Diagnosis Difficulty
    \item Other (Please specify)
\end{itemize}

\noindent \textit{Note: The workflow section conditionally branched to ask how practitioners resolved the specific challenges they selected in Q5, alongside their satisfaction levels. Below represents the core sequence.}

\noindent \textbf{Q6. How do you primarily resolve Diagnostic Uncertainty / Treatment Planning currently?}
\begin{itemize}[label={[ ]}, itemsep=0cm]
    \item Consult Seniors / Colleagues
    \item Post in Professional Social Media Groups (e.g., WhatsApp, Facebook)
    \item Use Standard Reference Tools / Books
    \item Online Search Engine
    \item Review Clinical Guidelines
    \item Other
\end{itemize}

\noindent \textbf{Q7. How satisfied are you with your current method for resolving Diagnostic/Treatment friction?}
\begin{itemize}[label={[ ]}, itemsep=0cm]
    \item Very Satisfied
    \item Satisfied
    \item Neutral
    \item Dissatisfied
\end{itemize}

\noindent \textbf{Q8. How do you currently calculate atopic dermatitis (AD) Severity Scores (e.g., EASI, SCORAD) in a busy clinic?}
\begin{itemize}[label={[ ]}, itemsep=0cm]
    \item Web/App-based Calculator
    \item Manual Calculation (Pen \& Paper)
    \item Visual Estimation (No formal scoring)
    \item Research setting only
    \item Not routinely performed in clinical practice
\end{itemize}

\noindent \textbf{Q9. How satisfied are you with your current method of AD Score Calculation?}
\begin{itemize}[label={[ ]}, itemsep=0cm]
    \item Very Satisfied
    \item Satisfied
    \item Neutral
    \item Dissatisfied
\end{itemize}

\noindent \textbf{Q10. What is your primary method for atopic dermatitis (AD) Patient Counselling?}
\begin{itemize}[label={[ ]}, itemsep=0cm]
    \item Verbal Discussion
    \item Printed Handouts
    \item Direct to Web Resources
    \item Refer to a Counsellor
\end{itemize}

\noindent \textbf{Q11. How satisfied are you with your current method of AD Patient Counselling?}
\begin{itemize}[label={[ ]}, itemsep=0cm]
    \item Very Satisfied
    \item Satisfied
    \item Neutral
    \item Dissatisfied
\end{itemize}

\subsubsection*{Section 4: Artificial Intelligence (AI) Adoption}
\noindent \textbf{Q12. Are you currently using any Artificial Intelligence (AI) tools in your clinical practice or academic workflow?}
\begin{itemize}[label={[ ]}, itemsep=0cm]
    \item Yes \textit{(Proceeds to Branch A)}
    \item No \textit{(Proceeds to Branch B)}
\end{itemize}

\vspace{0.2cm}
\noindent \textbf{Branch A: For AI Users}

\noindent \textbf{Q13a. Which AI tools do you primarily use? (Select all that apply)}
\begin{itemize}[label={[ ]}, itemsep=0cm]
    \item General LLMs (e.g., ChatGPT, Gemini, Claude)
    \item Specialized Diagnostic / Image Analysis Apps
    \item Other (Please specify)
\end{itemize}

\noindent \textbf{Q14a. What is your primary purpose for using these AI tools? (Select all that apply)}
\begin{itemize}[label={[ ]}, itemsep=0cm]
    \item Summarizing Research / Literature
    \item Patient Education / Drafting Handouts
    \item Drafting Clinical Notes
    \item Academic Work / Structuring Publications
    \item Analyzing Clinical Images
    \item Treatment Planning Assistance
    \item Other
\end{itemize}

\noindent \textbf{Q15a. What is the primary advantage you have noticed from using AI?}
\begin{itemize}[label={[ ]}, itemsep=0cm]
    \item Time Savings
    \item Easier Access to Information
    \item Treatment Planning Assistance
    \item Diagnostic Confidence
    \item Other
\end{itemize}

\noindent \textbf{Q16a. What is your overall satisfaction with the AI tools you currently use?}
\begin{itemize}[label={[ ]}, itemsep=0cm]
    \item Very Satisfied
    \item Satisfied
    \item Neutral
    \item Dissatisfied
\end{itemize}

\vspace{0.2cm}
\noindent \textbf{Branch B: For Non-AI Users}

\noindent \textbf{Q13b. What is the primary reason you do not currently use AI tools?}
\begin{itemize}[label={[ ]}, itemsep=0cm]
    \item Never felt the need
    \item Cost of software/tools
    \item Lack of training / Unsure how to use them
    \item No time to learn new tools
    \item Privacy concerns
    \item Tried a few but was not impressed / Lack of clinical utility
    \item Fear of errors / Misdiagnosis
    \item Other (Please specify)
\end{itemize}

\subsubsection*{Section 5: Ethical Concerns \& Future Promise}
\noindent \textit{(Displayed to all respondents)}

\vspace{0.2cm}
\noindent \textbf{Q17. What are your primary concerns regarding the integration of AI in dermatology? (Select all that apply)}
\begin{itemize}[label={[ ]}, itemsep=0cm]
    \item Patient Anxiety / Self-Misdiagnosis (Patients using AI at home)
    \item Non-Derm Usage / Competition (General Practitioners or Pharmacists using AI to treat skin conditions)
    \item Algorithmic Bias (Poor accuracy on diverse Indian skin tones)
    \item Misdiagnosis / Errors by the AI
    \item Deskilling / Over-reliance (Loss of clinical acumen)
    \item Erosion of the Patient-Doctor Relationship
    \item Data Privacy / Security Breaches
    \item Other (Please specify)
\end{itemize}

\noindent \textbf{Q18. Where do you see the most promising future application of Artificial Intelligence in dermatology?}
\begin{itemize}[label={[ ]}, itemsep=0cm]
    \item Medical Education
    \item Disease Registries \& Epidemiology
    \item Administrative Workflow / Note Drafting
    \item Advanced Image Diagnostics
    \item Other
\end{itemize}

\vspace{0.2cm}
\noindent \textbf{Q19. Do you have any additional comments regarding AI or your daily clinical challenges? (Optional)} \\
\textit{[ Open Text Field ]}

\end{document}